\documentclass[journal, 10pt, twocolumn]{IEEEtran}
\usepackage[utf8]{inputenc}
\usepackage{amsmath}

\usepackage{amsfonts}
\usepackage{amssymb}
\usepackage{cite}
\usepackage{tabularx}
\usepackage{booktabs}
\usepackage{multirow}
\usepackage{cases}
\usepackage{empheq}
\usepackage{comment}
\usepackage{relsize}
\usepackage{graphicx}
\usepackage{subcaption}
\usepackage{lineno}
\usepackage{xcolor}

\graphicspath{ {Figure/} }
\title{Full-duplex UWA communication with \\a two-element transducer}
\author{Lu Shen,~\IEEEmembership{Member,~IEEE}, Benjamin Henson,~\IEEEmembership{Member,~IEEE}, \\Long Shi,~\IEEEmembership{Member,~IEEE}, Yuriy Zakharov,~\IEEEmembership{Senior Member,~IEEE}
}

\begin{document}
\maketitle
\begin{abstract} 
In this work we present a full-duplex (FD) underwater acoustic (UWA) communication system simultaneously transmitting and receiving acoustic signals in the same frequency bandwidth. To simplify the FD hardware, the system exploits a recently designed two-element transducer capable of simultaneously transmitting and receiving signals. The key challenge of implementing an FD system is to cancel the strong self-interference (SI) from the near-end transmitter. By using advanced adaptive filtering algorithms providing high accuracy channel estimates, a high level of SI cancellation can be achieved when the far-end signal is absent. However, the SI channel estimation performance is limited in FD scenarios since the far-end signal acts as an interference. In this paper, we propose an FD UWA communication system which alternates between the SI cancellation and far-end data demodulation. An adaptive Rake combiner with multipath interference cancellation is implemented to improve the demodulation performance in time-varying multipath channels. The performance of the FD UWA system is evaluated in lake experiments. The proposed adaptive Rake combiner with multipath interference cancellation significantly outperforms the conventional Rake combiner in all the experiments. With the new Rake combiner, the detection performance of the proposed FD UWA system is comparable with that of the half-duplex system. 
\end{abstract}


\section{Introduction}
Underwater acoustic (UWA) communication suffers from limited frequency bandwidth for data transmission~\cite{stojanovic2009underwater}. To increase the spectral efficiency within the available bandwidth, we consider the full-duplex (FD) operation, which allows a simultaneous transmission and reception of acoustic signals in the same frequency bandwidth~\cite{choi2010achieving, duarte2010full, jain2011practical, duarte2012experiment}.
Apart from the far-end signal, the near-end receiver also receives a strong self-interference (SI) signal from the near-end transmitter. 
Therefore, a high level of SI cancellation (SIC) is required to detect the weak far-end signal.

In FD UWA systems, the spatial-domain SIC can be used for the SI suppression. In~\cite{hsieh2023full}, a spatial SIC scheme is proposed, which uses a phased array of transducers and acoustic vector sensors. The SIC is achieved by adjusting the beampattern of the phased array based on the direction of arrivals estimated by the acoustic vector sensors. Sea experimental results demonstrate a spatial SIC of 40 to 60~dB depending on the steering angle of the near-end transducer.
In~\cite{lu2022spatial}, an FD UWA system with a combination of spatial SIC and digital SIC is proposed and investigated in an anechoic pool; this system requires the knowledge of the direction of the far-end transmitter, which is not always available in practice. 
Recently, an FD UWA system has been proposed that performs a joint SI cancellation and far-end signal equalization~\cite{towliat2024joint}. This system is capable of operating at a distance of 200 m at a carrier frequency of 10 kHz. However, it requires a hydrophone array and a spatial separation of more than 7~m between the near-end projector and hydrophone arrays; this system cannot demonstrate reasonable performance with a single hydrophone.
UWA communication systems with antenna arrays are difficult for practical deployment. However, even without using an antenna array, in an FD system, typically the transmitting and receiving transducers are two separate devices. 
It can be argued~\cite{zhou2015full} that using two-element antenna arrays at the near-end and far-end can double the data rate of a half-duplex (HD) system compared to a single element HD system, which would match the ideal FD performance, and thus making the complicated FD designs unnecessary. Such a MIMO (multi-input multi-output) communication would however require a substantial space separation of the antenna elements to make the four MIMO channels statistically independent.
In~\cite{henson2022full}, we proposed an FD transducer that can be used for simultaneous transmission and reception of acoustic signals. 
The FD transducer contains two piezo-ceramic cylinders. One cylinder acts as a hydrophone and the other a projector, and the distance between centres of two cylinders is 23~mm.
We use this transducer in our UWA FD experiments.         

For FD terrestrial radio communications, a combination of analogue and digital cancellation is normally used. Analogue cancellation is applied before digital cancellation to avoid the analogue-to-digital converter (ADC) saturation~\cite{choi2010achieving, duarte2010full, jain2011practical}. 
In FD UWA systems, compared to terrestrial radio systems, low-frequency signals are transmitted. In such a case, high resolution ADCs up to 24 bits can be used, which makes it feasible to achieve a high level of cancellation in the digital domain alone.
Therefore, digital cancellation is considered as the main SIC approach in FD UWA systems~\cite{qiao2018digital, qiao2018self, shen2020adaptive, shen2022bem, tsimenidisband}.
In this work, we implement and investigate the performance of an FD UWA system with digital cancellation.

Two factors limiting the digital SIC performance in FD UWA systems are  nonlinearities introduced by the equipment~\cite{qiao2018digital, shen2019digital, shen2020eq} and fast variation of the SI channel~\cite{shen2020adaptive, shen2022bem}. 
The dominant source of the nonlinearities is the power amplifier (PA). It is found that the SIC performance can be significantly improved by using the PA output as the reference signal (regressor) for SI channel estimation compared to the case of using the original digital data~\cite{li2011full, shen2019digital}. 
After addressing the nonlinearities caused by the PA, a high level of digital SIC can be achieved in time-invariant scenarios with a classical recursive least-squares (RLS) adaptive filter~\cite{haykin2002adaptive}.
In practice, the SI channel can be fast time-varying, e.g., due to reflections from the moving sea surface~\cite{li2015interference}. 
To achieve a high level of digital SIC in time-varying channels, advanced adaptive filtering algorithms with good tracking performance have been recently proposed~\cite{shen2020adaptive, shen2022bem, niedzwiecki2022adaptive, niedzwiecki2023bidirectional}. 
The SIC performance provided by these adaptive algorithms has been evaluated in shallow lake experiments. However, in those experiments, the far-end transmission is not considered and the SIC performance is only limited by the ambient noise level. 

With the FD operation, the received signal includes the near-end SI, background noise, and also the far-end signal. When estimating the near-end SI channel, the far-end signal acts as an additional interference (noise), which reduces the overall SI to noise ratio.  To achieve a high level of SIC, the influence of the far-end signal, in addition to the ambient noise, on the SIC performance must be taken into account. 

To avoid the impact of the far-end signal on the near-end SI channel estimation performance, a scheme is proposed in~\cite{qiao2018digital} with estimation of the near-end SI channel before the far-end signal arrives. An over-parametrization based RLS algorithm with a sparsity constraint is proposed for the SI channel estimation and evaluated in a water tank with a static multipath channel.
This approach works well in time-invariant scenarios. However, the SI channel needs to be frequently re-estimated to ensure a good tracking performance in fast time-varying scenarios. \textcolor{black}{In such a case, it is difficult to guarantee sufficiently long time intervals for the SI channel estimation, non-overlapping with the far-end transmission, which is required in this system.} 


\textcolor{black}{To reduce the impact of the far-end signal on the SIC performance, we propose an FD UWA system which alternates between the near-end SIC and far-end data demodulation in turbo iterations. The proposed system does not require a non-overlapping period between the near-end and far-end transmissions. 
In the first iteration, the SI channel is estimated, treating the far-end signal as an extra noise. 
At the receiver, the residual signal after SIC is used for tentative far-end data demodulation.
In the second iteration, the far-end signal is reconstructed using the far-end data and channel estimates and then removed from the received signal. 
After that, a similar procedure is performed as in the first iteration. 
As the estimation performance for both the near-end SI channel and far-end channel is  improved in the second iteration, a better demodulation performance can be achieved. Further iterations can be useful to improve the communication performance. }

In~\cite{shen2022full}, the conventional Rake combiner~\cite{proakis2008digital} is used for demodulation of far-end data in multipath channels. 
The Rake combiner provides the maximal ratio combining for signal components propagated over various channel paths. However, such a combiner ignores the multipath interference, i.e., the interference from one multipath signal component to another, which significantly degrades the receiver performance.  
In~\cite{fathi2006hybrid}, a receiver with linear equalization, multipath interference reconstruction and interference cancellation is proposed. Simulation results indicate that the receiver performance is significantly improved due to the multipath interference cancellation. 
In~\cite{shen2023rake}, \textcolor{black}{a Rake combiner with multipath interference cancellation (Rake-IC combiner)} is implemented in a UWA modem with half-duplex data packet transmission. This Rake-IC combiner shows good demodulation performance in static/slow time-varying scenarios. 
\textcolor{black}{In this paper, we propose an adaptive version of the Rake-IC combiner, which is capable of providing good demodulation performance in FD time-varying scenarios. }

\textcolor{black}{To evaluate the performance of the FD UWA system, both half-duplex and FD communication experiments are conducted in a lake using recently developed two-element transducers~\cite{henson2022full}. 
Low signal-to-interference ratios (SIRs) are considered in the FD experiments to demonstrate that the proposed FD system is capable of operating at high SI levels.}
The proposed Rake-IC combiner outperforms the conventional Rake combiner in all the experiments. 
Furthermore, results indicate that the proposed FD system with the Rake-IC combiner can provide the demodulation performance close to that of the half-duplex system. 
The contributions of this paper are as follow.
\begin{itemize}
\item An FD UWA system with multiple iterations is proposed for cancellation of the interference from the far-end transmission when estimating the SI channel and cancellation of the SI when demodulating the far-end data.  
\item An adaptive version of the Rake-IC combiner is proposed to improve the detection performance in time-varying multipath channels.
Its performance is investigated in lake experiments.
\item \textcolor{black}{Lake experiments are conducted to evaluate the proposed FD UWA system using two-element transducers.} In all the experiments, the detection performance of the proposed FD UWA system is comparable with that of the half-duplex system.
\end{itemize}

The paper is organized as follows. In Section~\ref{sec:system_model}, the FD UWA system structure is presented. Section~\ref{sec:receiver} describes the Rake-IC combiner. Section~\ref{sec:lake_exp} investigates the performance of the FD UWA system and the proposed receiver in lake experiments. Section~\ref{sec:conclude} concludes the paper.
\textcolor{black}{In Appendix, we provide a general description of the adaptive filters used in the proposed FD receiver.}

\begin{figure}
\centering
 \includegraphics[width=0.49\textwidth]{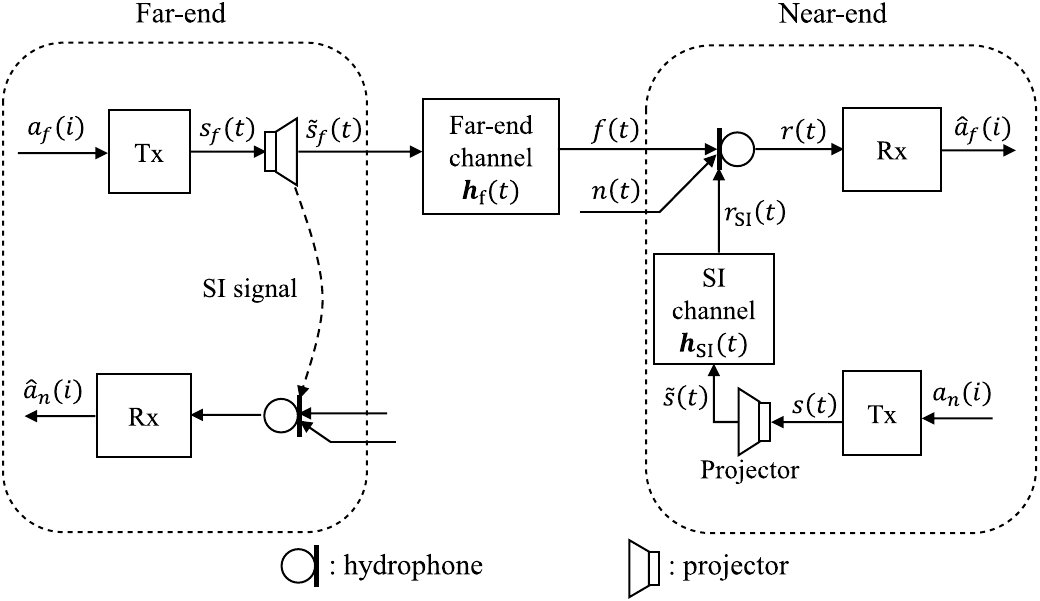}      
 \caption{Block diagram of the FD UWA system.\label{Fig:blockDiagram} The sample index with the symbol rate $f_d$ is denoted by $i$; $a_n(i)$ and $a_f(i)$ represent the near-end and far-end data symbols, respectively; $s(t)$ and $s_f(t) $ denote the analogue PA outputs of the near-end and far-end transmitters; $\tilde{s}(t)$ and $\tilde{s}_f(t)$ represent the near-end and far-end projector outputs; $r(t)$ denotes the received signal.}
\end{figure}
\section{System model}\label{sec:system_model}
The block diagram of the FD UWA system is shown in Fig.~\ref{Fig:blockDiagram}. 

The signal at the near-end receiver is given by:
\begin{align}
\begin{split}
r(t) &= \tilde{s}(t) * h_\mathrm{SI}(t) + \tilde{s}_f(t) * h_\mathrm{f}(t) + n(t) \\
     &=  r_\mathrm{SI}(t) + f(t) + n(t),
     \end{split}
\end{align}
where $*$ denotes the convolution, $r_\mathrm{SI}(t) = \tilde{s}(t) * h_\mathrm{SI}(t)$ is the SI, $\tilde{s}(t)$ is the signal emitted by the near-end projector, $h_\mathrm{SI}(t)$ represents the SI channel impulse response, $f(t) = \tilde{s}_f(t) * h_\mathrm{f}(t)$ is the far-end signal, $\tilde{s}_f(t)$ is the signal emitted by the far-end projector, $h_\mathrm{f}(t)$ represents the far-end channel impulse response and $n(t)$ is the background noise. The near-end and far-end channel responses, $h_\mathrm{SI}(t)$ and $h_\mathrm{f}(t)$, and the far-end signal $f(t)$ are unknown at the receiver and should be estimated. 

In subsection~\ref{subsec:Tx_signal}, description of signals used for near-end and far-end transmission is given. In subsection~\ref{subsec:Rx}, the receiver structure is introduced.

\subsection{Transmitted signals}\label{subsec:Tx_signal}
\begin{figure}
\centering
 \includegraphics[width=0.5\textwidth]{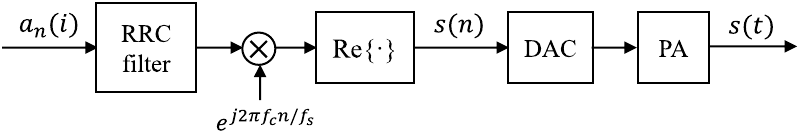}      
        \caption{Near-end transmitter. The sample index with the symbol rate $f_d$ and sampling rate $f_s$ are denoted by $i$ and $n$, respectively; $a_n(i)$ represents the near-end data symbols;  $s(n)$ denotes the passband digital signal; $s(t)$ denotes the analogue PA output.\label{Fig:Tx}}
\end{figure}

The structure of the near-end transmitter is shown in Fig.~\ref{Fig:Tx}.
For the near-end transmission, a pseudo-random sequence of Binary Phase Shift Keying (BPSK) data symbols $a_n(i)$ is used. The data symbols are up-sampled and pulse-shaped by a root-raised cosine (RRC) filter. The RRC filter output is then up-converted to the carrier frequency $f_c$. 
Afterwards, the signal is digital-to-analogue converted, amplified by a PA, and the near-end PA output $s(t)$ is transmitted by a transducer. 

For the far-end transmission, Quadrature Phase Shift Keying (QPSK) symbols consisting of superimposed binary pilot and data symbols~\cite{alameda2005synchronisation, zakharov2014ofdm} are used:
\begin{equation}
a_f(i) = p(i) + jd(i),
\end{equation}
where $p(i)$ is a binary pseudo-random pilot sequence and $d(i)$ is the information data symbols obtained by encoding transmitted data using a convolutional code and interleaving the encoded symbols.
The far-end symbols $a_f(i)$ go through the same modulation process as shown in Fig.~\ref{Fig:Tx}. 

\subsection{Receiver structure}\label{subsec:Rx}
The receiver performs two tasks, the first task is the near-end SIC, and the second task is the far-end data demodulation. 

\begin{figure}
\centering
 \includegraphics[width=0.4\textwidth]{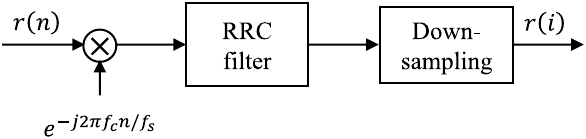}      
 \caption{Complex demodulation. $r(n)$ and $r(i)$ represent the passband and baseband received signals, respectively.\label{Fig:front-end}} 
\end{figure}
The passband received signal $r(t)$ is first \textcolor{black}{converted by an analogue-to-digital converter (ADC)} to produce the passband samples $r(n)$, which then go through the complex demodulation process to produce the baseband samples $r(i)$ as shown in Fig.~\ref{Fig:front-end}, where an RRC filter is used for the low-pass filtering. 

\begin{figure*}
\centering
 \includegraphics[width=1\textwidth]{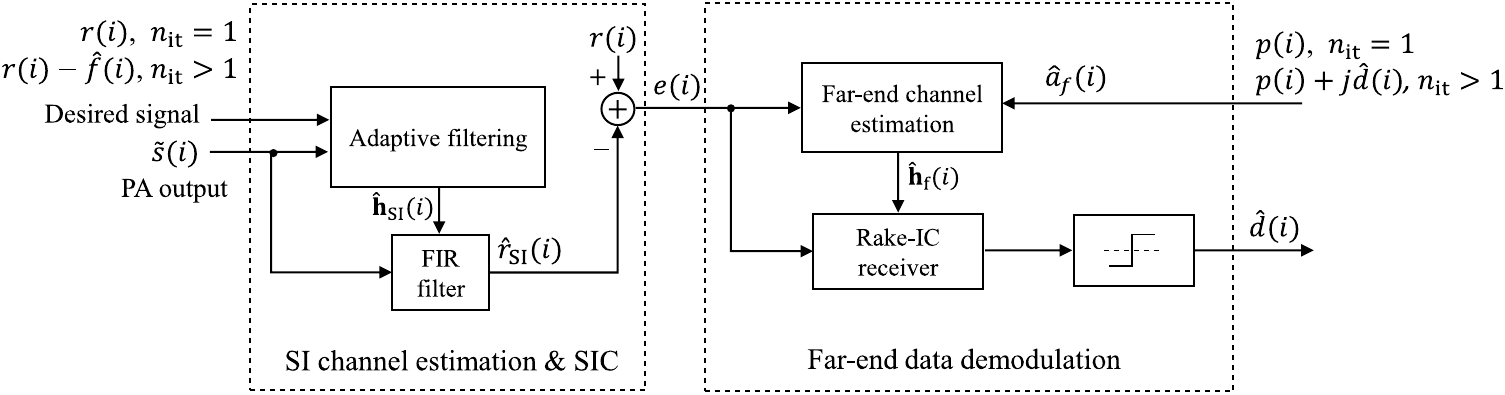}      
 \caption{Receiver structure. $n_{it}$ is the iteration number, $r(i)$ is the baseband received signal, $e(i)$ is the residual signal after SIC and $\hat{f}(i)$ is the estimate of the far-end signal.     \label{Fig:Rx}} 
\end{figure*}

The general structure of the receiver is shown in Fig.~\ref{Fig:Rx}.
In the first iteration ($n_\text{it} = 1$), since the far-end signal estimate $\hat{f}(i)$ is not available yet, the received signal $r(i)$ is the desired signal of an adaptive filter used for the SI channel estimation and near-end SIC. To prevent the PA nonlinearity affecting the SI channel estimation, the baseband digitalized PA output $\tilde{s}(i)$ is used as the regressor in the adaptive filter, instead of using the symbols $a_n(i)$~\cite{shen2019digital}. The baseband signal $\tilde{s}(i)$ is obtained through the same complex demodulation process as shown in Fig.~\ref{Fig:front-end}. 

The baseband PA output $\tilde{s}(i)$ is fed to a finite impulse response (FIR) filter whose impulse response is a delayed copy~\cite{shen2020adaptive, shen2022bem} of the impulse response of the adaptive filter, i.e., the SI channel estimate $\hat{\mathbf{h}}_\mathrm{SI}(i)$. To perform the SIC, the FIR filter output $\hat{r}_\mathrm{SI}(i)$ is subtracted from the received signal $r(i)$. The residual signal $e(i)$ is an estimate of the far-end signal (plus noise) which is used for the far-end channel estimation and far-end data demodulation. 

\begin{figure*}
\begin{subfigure}{0.38\textwidth}
\centering
 \includegraphics[width=\textwidth]{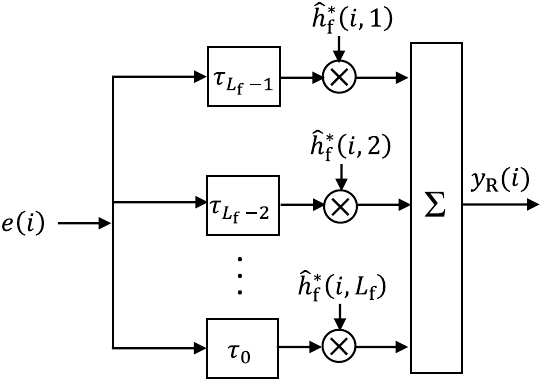}      
  \caption{\label{Fig:MRC} Conventional Rake combiner.}
  \end{subfigure}
  \hfill
  \begin{subfigure}{0.58\textwidth}
  \centering
  \includegraphics[width=\textwidth]{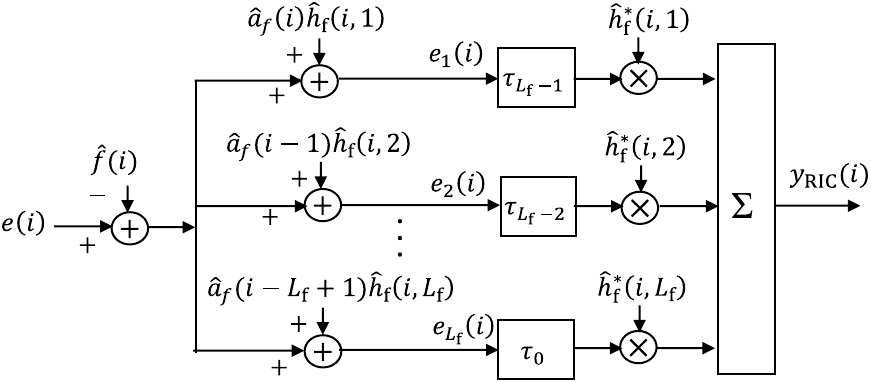}      
  \caption{\label{Fig:MRC_with_IC} Rake-IC combiner.}
  \end{subfigure}
  \caption{Block diagrams of the Rake combiners; $e(i)$ is the residual signal after SIC; $\hat{a}_f(i)$ is the far-end data estimate; $e_l(i)$ is the recovered $l$th multipath signal component; $\tau_l$ represents a delay of $l$ symbol intervals; $y_\text{R}(i)$ and $y_\text{RIC}(i)$ are the output signals of the conventional Rake combiner and the Rake-IC combiner, respectively.}
\end{figure*}

The far-end channel estimate $\hat{\mathbf{h}}_\mathrm{f}(i)$ is obtained by another adaptive filter. An estimate $\hat{a}_f(i)$ of the far-end data $a_f(i)$, obtained at the previous iteration, and the residual signal $e(i)$ are used as the regressor and the desired signal of the adaptive filter, respectively. 
In the first iteration ($n_\text{it}~=~1$), there is no estimate of the far-end data $d(i)$, but the pilot symbols $p(i)$ are known and therefore we use $\hat{a}_f(i) = p(i)$. In further iterations ($n_\text{it}~>~1$), there is a tentative estimate $\hat{d}(i)$ of the far-end data, and therefore we use an improved estimate $\hat{a}_f(i) = p(i) + j \hat{d}(i)$. 

In this paper, two types of Rake combiner are considered for the far-end data demodulation, the conventional Rake combiner and Rake-IC combiner with multipath interference cancellation; they will be described in Section~\ref{sec:receiver}. 
The imaginary part of the Rake combiner output is interleaved and decoded using the Viterbi decoder~\cite{proakis2008digital}. The decoded message is encoded and interleaved in the same way as it is done in the transmitter and thus the obtained sequence of symbols $\hat{d}(i)$ is used as an estimate of the far-end symbols $d(i)$.  

With $\hat{a}_f(i)$ and $\hat{\mathbf{h}}_\mathrm{f}(i)$ now available, an estimate  $\hat{f}(i)$ of the far-end signal $f(i)$ is reconstructed by filtering $\hat{a}_f(i)$ in a FIR filter with the impulse response $\hat{\mathbf{h}}_\mathrm{f}(i)$; this estimate will be used in the next iteration.

\textcolor{black}{If the SI channel and far-end channel are estimated separately (not jointly), the residual SI level after the first SIC iteration depends on the level of the combined far-end signal and background noise. Practically, with high performance SI channel estimators, the SI can be cancelled to this combined level.
If the far-end signal level is higher than the noise level, which is a scenario of interest here, such a cancellation can provide the far-end signal-to-interference-plus-noise ratio (SINR) lower than 0 dB. Such a low SINR makes the demodulation of the far-end signal a challenge.
The use of low-rate codes can improve the demodulation performance, which in turn improves the estimate $\hat{f}(i)$. In the next iteration, with a more accurate SI signal estimate $r(i) - \hat{f}(i)$, a better SI channel estimate is achieved, and so on for subsequent iterations.
However, the use of low-rate convolutional codes results in a low spectral efficiency of the FD communication. There is a trade-off between the performance of the FD system and its spectral efficiency.
In Section~\ref{sec:lake_exp}, we investigate the performance of the HD and FD systems with different convolutional codes listed in Table~\ref{Tab:code}. 
}

\section{Rake combiner with interference cancellation}\label{sec:receiver}

The conventional Rake combiner shown in Fig.~\ref{Fig:MRC} is implemented as an FIR filter with its $k$th filter coefficient at the $i$th time instant given by:
\begin{equation}
\hat{h}_\mathrm{R}(i, k) = {\hat{h}}^*_\mathrm{f}(i, L_\mathrm{f}-k),
\end{equation}
where $\hat{\mathbf{h}}_\mathrm{f}(i)$ is a far-end channel estimate at the $i$th time instant (a vector of size $L_\text{f} \times 1$ with channel tap estimates) and $L_\mathrm{f}$ is the length of the far-end channel impulse response. The conventional Rake combiner performs the maximal-ratio combining but it ignores the multipath interference, while the Rake-IC combiner performs the multipath interference cancellation before the maximal-ratio combining. 

In this section, we introduce an adaptive version of the Rake-IC combiner proposed in~\cite{shen2023rake}, which is implemented assuming that the multipath channel is static. The adaptive Rake-IC combiner can deal with time-varying multipath interference cancellation.
The block diagram of the proposed Rake-IC combiner is shown in Fig.~\ref{Fig:MRC_with_IC}. 

\textcolor{black}{In each branch in Fig.~\ref{Fig:MRC_with_IC}, only one multipath signal component is kept for processing. The other multipath components are removed from the residual signal $e(i)$ based on the far-end channel estimate $\hat{\mathbf{h}}_\mathrm{f}(i)$. This process can either be individually performed for each branch, or it can be performed in a more computationally efficient way as shown in Fig.~\ref{Fig:MRC_with_IC}. First, all of the signal components are removed from $e(i)$, and then every signal component is added to the difference $e(i)-\hat{f}(i)$ individually on each branch with a corresponding delay. In such a case, we obtain the $l$th multipath signal $e_l(i)$ free from the multipath interference:
\begin{equation}
e_l(i) = e(i)-\hat{f}(i) + \hat{a}_f(i)\hat{h}_f(i, l).
\end{equation}
Then, the maximal-ratio combining results in the signal
\begin{equation}
y_\text{RIC}(i) = \sum_{l = 1}^{L_\mathrm{f}}e_l(i-L_\mathrm{f}+l)\hat{h}^*_\mathrm{f}(i,l).
\end{equation}}


In Section~\ref{sec:lake_exp}, we show that the demodulation performance can be significantly improved when using the proposed Rake combiner instead of the conventional Rake combiner.

\section{Lake experiments}\label{sec:lake_exp}
In this section, we investigate the performance of the FD UWA system in two experimental sites. For comparison, HD experiments were also conducted, i.e., experiments with the far-end transmission only.

\begin{figure}
\centering
   \includegraphics[width=0.4\textwidth]{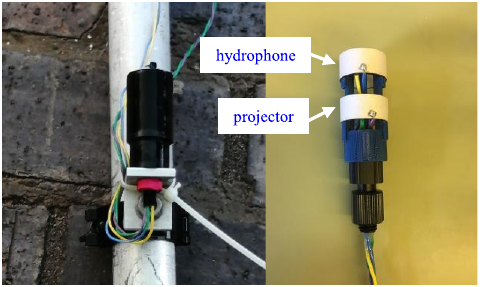}      
  \caption{\label{Fig:FD_tx} A two-sensor transducer used in the lake experiments.}
\end{figure}

The lake experiments were conducted using the recently developed two-sensor FD transducer (see Fig.~\ref{Fig:FD_tx}) which contains two piezo-ceramic cylinders, one of them is used as a projector and the other one is used as a hydrophone~\cite{henson2022full}. 

For both the near-end and far-end transmission, the transmitted signal is sampled at $f_s = 192$~kHz. The RRC filter with a roll-off factor of  $0.2$ is used for pulse-shaping and low-pass filtering. In every experiment, signals of 60~s length are transmitted. A silence period of 5~s is added at the beginning of the transmitted signal to measure the background noise. Far-end data are encoded with convolutional codes of different code rates as detailed in Table~\ref{Tab:code}.

\textcolor{black}{A Zoom F4 recorder is used for recording received signals from the hydrophones~\cite{zoomF4}. The ADC in the Zoom F4 recorder has 24-bit resolution.}

\begin{figure}
\centering
   \includegraphics[width=0.4\textwidth]{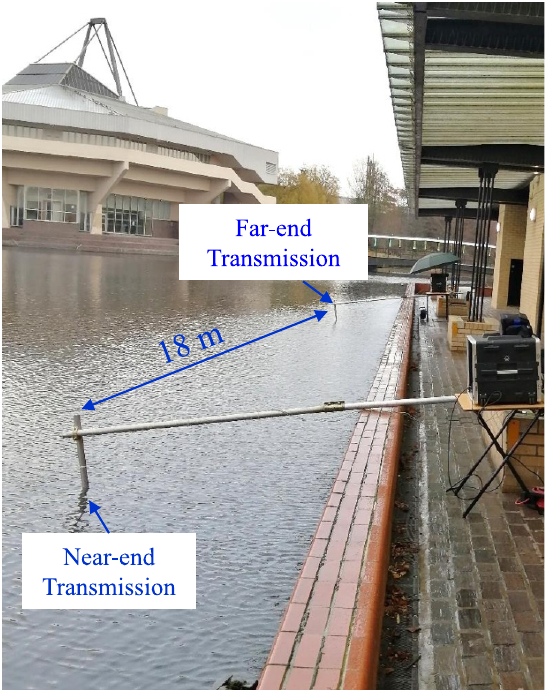}      
  \caption{\label{Fig:FD_exp} Setup of the FD and HD experiments in the University of York.}
\end{figure}
\begin{table}[]
\caption{List of convolutional codes~\cite{proakis2008digital} used in the experiments}\label{Tab:code}
\centering
\begin{tabular}{|l|l|l|}
\hline
Code Rate & \multicolumn{1}{c|}{Constraint Length} & \multicolumn{1}{c|}{Polynomials in Octal} \\ \hline
1/4       & 8                                      & {[}235 275 313 357{]}                        \\ \hline
1/3       & 8                                      & {[}225 331 367{]}                        \\ \hline
1/2       & 6                                      & {[}53 75{]}                              \\ \hline
2/3       & {[}5 4{]}                              & {[}23 35 0; 0 5 13{]}                    \\ \hline
\end{tabular}
\end{table}

\subsection{Experimental site 1}\label{subsec:campus_lake}

\subsubsection{Experimental setup}\label{subsec:exp_setup}
The first experimental site is the shallow lake at the University of York. The maximum depth of the experimental site is around 1~m. The experimental setup is shown in Fig.~\ref{Fig:FD_exp}. Both transducers were placed at approximately 0.5~m depth and the distance between them was around 18~m. 

Both near-end and far-end signals are transmitted at $f_c = 36$~kHz carrier frequency with $4$~kHz bandwidth. \textcolor{black}{The source level for the near-end projector is 161~dB re $\mu$Pa $@1$~m. The source level for the far-end projector is 135~dB re $\mu$Pa $@1$~m. A lower source level is used for the far-end projector to obtain a low far-end SNR. After the front-end processing,}
the far-end SNR was around $16$~dB. For the FD experiment, the received SI to noise ratio was around 67~dB, i.e., the \textcolor{black}{signal-to-interference ratio is about -51~dB.}

\subsubsection{SI and far-end channel estimation}\label{subsec:ch_est}
The SI channel estimation is significantly more demanding than the far-end channel estimation, since the level of the required SI cancellation is much higher than that of the far-end signal cancellation. The SI and far-end channels in this experiment were fast-varying due to the transducers positioned close to the lake surface. Therefore, for the SI channel estimation, we use the homotopy SRLS-L-DCD (HSRLS-L-DCD) adaptive filter~\cite{shen2022bem}. This adaptive filter uses the basis function expansion of the channel variations to improve its tracking performance and exploits the channel sparsity to improve the estimation accuracy. For the far-end channel estimation, the delayed sliding-window RLS (SRLSd) adaptive filter~\cite{shen2020adaptive} is used. The HSRLS-L-DCD algorithm could also be used for the far-end channel estimation, but the SRLSd algorithm is less complicated. \textcolor{black}{Detailed description of the SRLSd and HSRLS-L-DCD algorithms are given in Appendix.}

%
%
%
%
%
%
%

The HSRLS-L-DCD adaptive filter length is $L_\text{SI} = 100$ taps, which corresponds to a delay spread of $25$~ms (the baseband sampling rate is $f_d$ = 4~kHz). The sliding window length in the first iteration is set to $M = 1401$ and, in further iterations, $M = 1001$. A high value of $M$ allows better performance in noisy environment, which is the case in the first iteration when the interference from the far-end signal has not been cancelled yet. A smaller $M$ allows better tracking of the channel variations in less noisy environment, which is the case in the second and further iterations after cancelling the far-end interference. The adaptive filter uses the parabolic approximation of the time variations of the complex multipath amplitudes~\cite{shen2022bem}.

The SRLSd adaptive filter length is $L_\mathrm{f} = 20$ taps (5~ms) and the sliding window length is $M = 1401$ in the first iteration and $M = 1001$ in the second and further iterations. The reasons for changing $M$ in the first and further iterations are similar to that for the HSRLS-L-DCD adaptive filter.


\begin{figure}
\centering
   \includegraphics[width=0.5\textwidth]{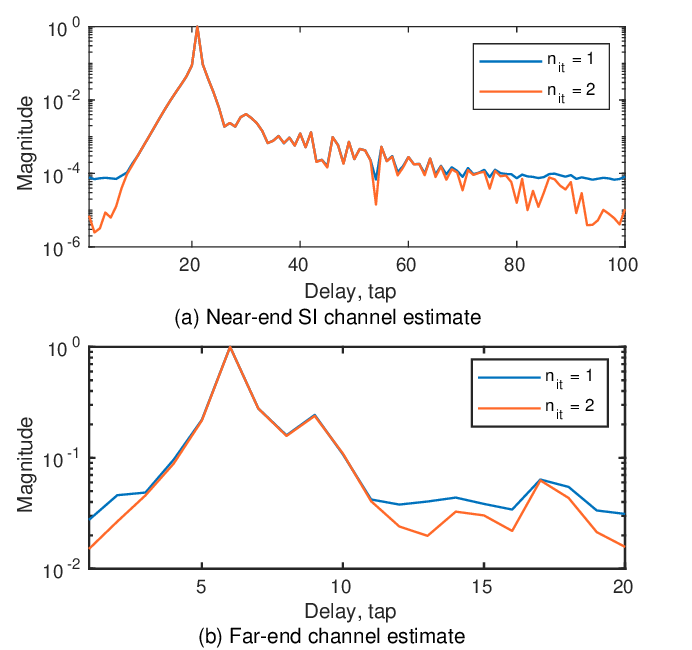}      
  \caption{\label{Fig:Near-end} Magnitude of the impulse response estimates of the SI channel and far-end channel in the FD experiment.}
\end{figure}

\begin{figure}
\centering
 \begin{subfigure}{0.5\textwidth}
         \centering
         \includegraphics[width=\textwidth]{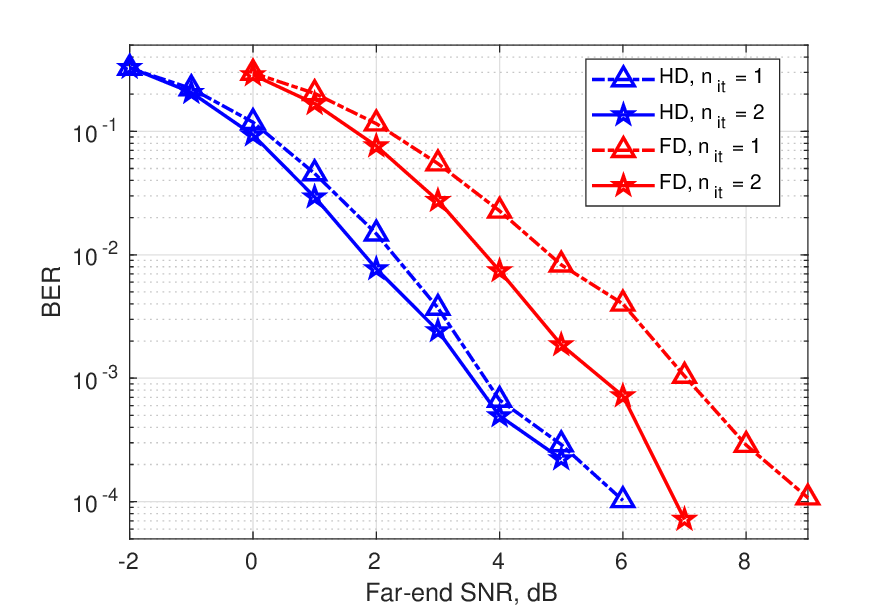}
         \caption{Conventional Rake combiner.\label{Fig:BER_1by4_Rake}}
 \end{subfigure}
\begin{subfigure}{0.5\textwidth}
         \centering
         \includegraphics[width=\textwidth]{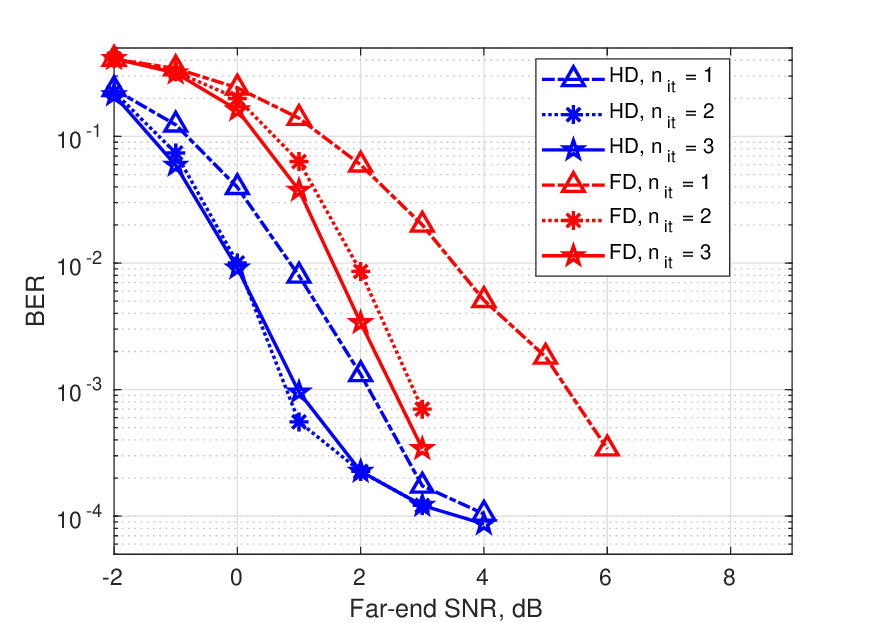}
         \caption{Rake-IC combiner.\label{Fig:BER_1by4_Rake-IC}}
 \end{subfigure}
\begin{subfigure}{0.5\textwidth}
         \centering
         \includegraphics[width=\textwidth]{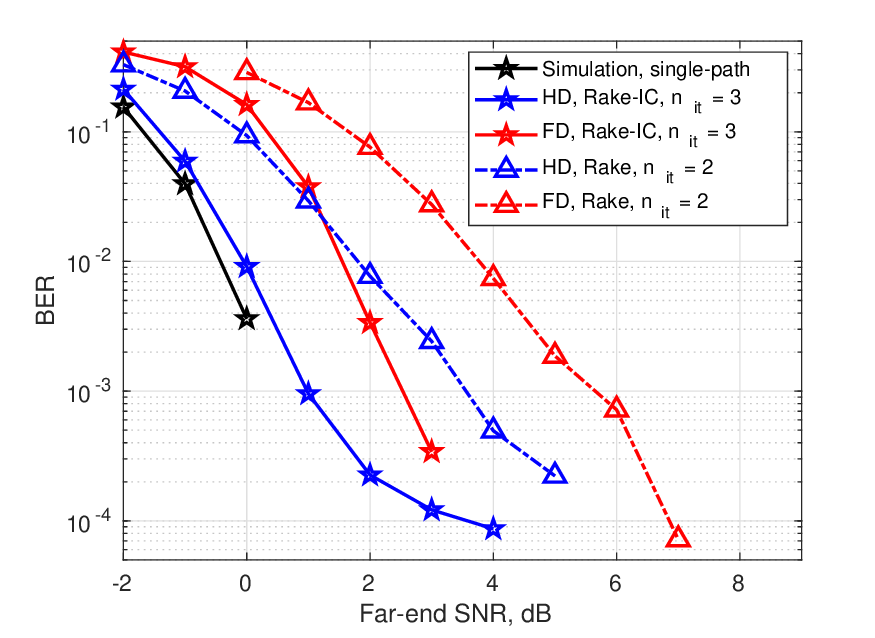}
         \caption{Rake and Rake-IC combiners.\label{Fig:BER_1by4_combined}}
 \end{subfigure}
  \caption{\label{Fig:BER} Experimental BER performance of HD and FD UWA systems with the conventional Rake and Rake-IC combiners with rate-1/4 convolutional code.}
\end{figure}

An example of the magnitude of the near-end and far-end channel impulse responses at the first and second iterations are shown in Fig.~\ref{Fig:Near-end}. 
In Fig.~\ref{Fig:Near-end}a, it can be seen that the multipath structure of the SI channel estimate is much clearer after the second iteration. The same is true for the far-end channel estimate shown in Fig.~\ref{Fig:Near-end}b. 
This demonstrates that both the SI and far-end channel estimation performance are improved at the second iteration.

\begin{figure}
\centering
 \begin{subfigure}{0.5\textwidth}
 \centering
   \includegraphics[width=\textwidth]{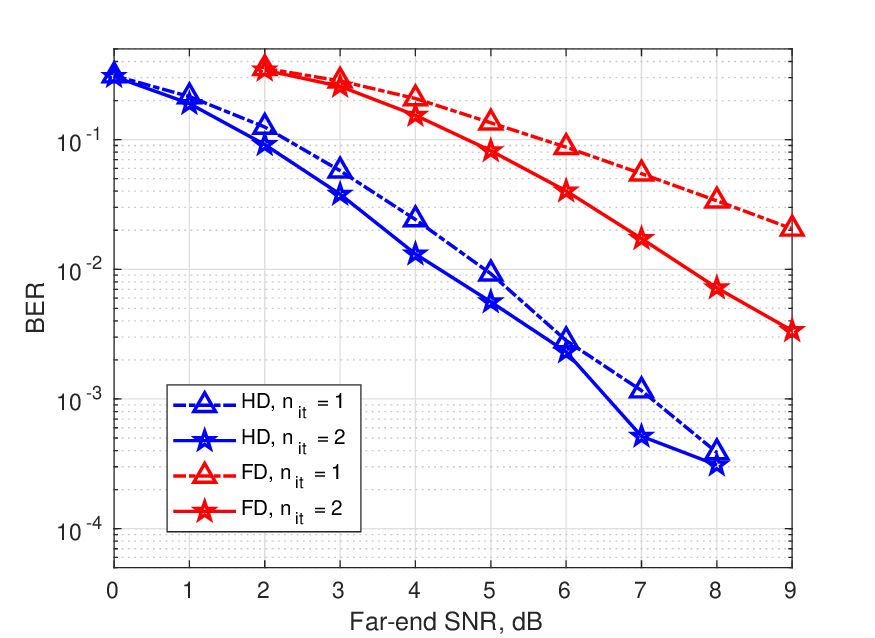}      
  \caption{Conventional Rake combiner.\label{Fig:BER_1by3_Rake}}
  \end{subfigure}
\begin{subfigure}{0.5\textwidth}
\centering
   \includegraphics[width=\textwidth]{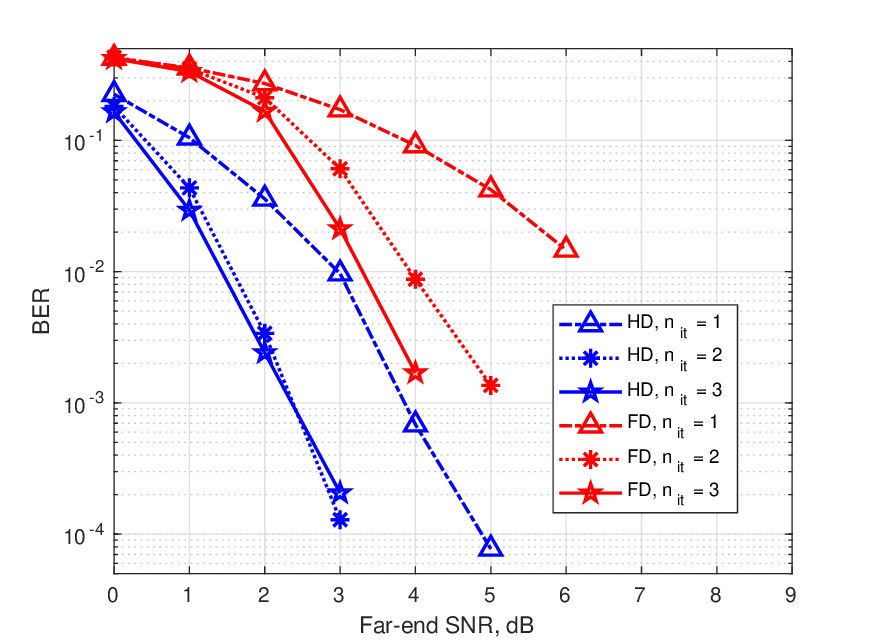}      
  \caption{Rake-IC combiner.\label{Fig:BER_1by3_Rake-IC}}
\end{subfigure}
\begin{subfigure}{0.5\textwidth}
\centering
   \includegraphics[width=\textwidth]{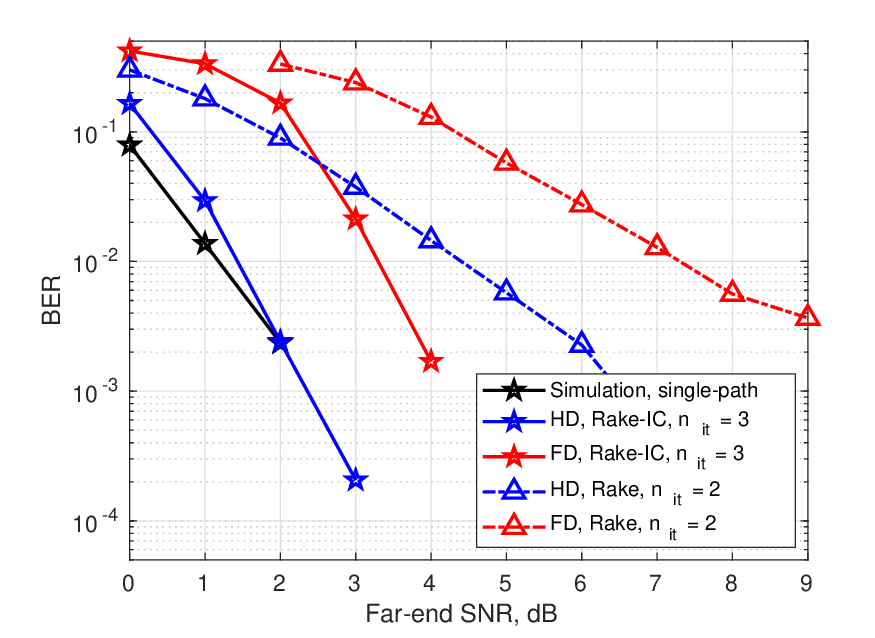}      
  \caption{Rake and Rake-IC combiners.\label{Fig:BER_1by3_combined}}
\end{subfigure}
\caption{Experimental BER performance of HD and FD UWA systems with the conventional Rake and Rake-IC combiners with rate-1/3 convolutional code.\label{Fig:BER_1by3}}
\end{figure}

\subsubsection{Demodulation performance}\label{subsec:demod}
To obtain the demodulation performance at various far-end SNRs,  complex-valued Gaussian random noise with variance $\sigma_n^2$ is added to the baseband received signal in the HD and FD experiments, where $\sigma_n^2$ is computed as: $\sigma_n^2 = (P_s-P_n)/\mathrm{SNR} - P_n$, 
and $P_s$ and $P_n$ are the average power of the baseband received signal and background noise, respectively, in the HD experiment. \textcolor{black}{The baseband received signal in the HD experiment includes the far-end signal and the background noise. The baseband received signal in the FD experiment includes the SI signal, the far-end signal and the background noise.}

We first consider results of experiments with a rate 1/4 convolutional code, presented in~Fig.~\ref{Fig:BER}. 
The BER performance of the conventional Rake combiner in HD and FD lake experiments is shown in Fig.~\ref{Fig:BER_1by4_Rake}. 
The demodulation performance is improved in the second iteration for both HD and FD experiments, with a performance gap of about 2~dB between these two transmission modes. In further iterations, the performance (not shown here) is the same.  

The BER performance of the Rake-IC combiner in HD and FD experiments is shown in Fig.~\ref{Fig:BER_1by4_Rake-IC}. Fig.~\ref{Fig:BER_1by4_Rake-IC} further demonstrates that using multiple iterations in the receiver is beneficial for the demodulation performance. In this case, the third iteration brings some extra improvement in the performance, and the performance gap between the FD and HD modes is somewhat smaller than~2~dB. 

For better comparison of the Rake combiners, Fig.~\ref{Fig:BER_1by4_combined} shows 
several BER curves selected from Fig.~\ref{Fig:BER_1by4_Rake} and Fig.~\ref{Fig:BER_1by4_Rake-IC}. Here, we also show the BER performance obtained with the same signals via numerical simulation for the single-path static HD channel; this is a benchmark indicating the best performance that can be achieved in the ideal conditions. It can be seen that the Rake-IC combiner outperforms the conventional Rake combiner by 2-3~dB, with higher improvement at lower BERs. With the Rake-IC combiner, the BER performance in the HD experiment approaches the single-path benchmark after two iterations. This shows that the Rake-IC combiner has achieved almost ideal (maximal-ratio) combining of the multipath components with complete cancellation of the multipath interference.


The demodulation performance of the Rake combiners with a rate 1/3 convolutional code are shown in Fig.~\ref{Fig:BER_1by3}. The BER curves in Fig.~\ref{Fig:BER_1by3_Rake} and~Fig.~\ref{Fig:BER_1by3_Rake-IC} further demonstrate the improvement in the demodulation performance achieved with turbo iterations; the performance improvement is especially high in the FD experiments. Selected BER curves for the Rake combiners in HD and FD experiments are shown in Fig.~\ref{Fig:BER_1by3_combined}. By comparing Fig.~\ref{Fig:BER_1by4_combined} and Fig.~\ref{Fig:BER_1by3_combined}, it can be seen that the SNR loss of the conventional Rake combiner to the single-path benchmark has significantly increased with the higher code rate. On the other hand, the performance of the Rake-IC combiner remains equally good for rate-1/3 and rate-1/4 convolutional codes. At a BER of $10^{-3}$, for the Rake-IC combiner, the SNR loss in the FD experiment is still about 2~dB compared to the HD experiment.

\subsection{Experimental site 2}\label{subsec:neptune_exp}
\begin{figure}
\centering
   \includegraphics[width=0.5\textwidth]{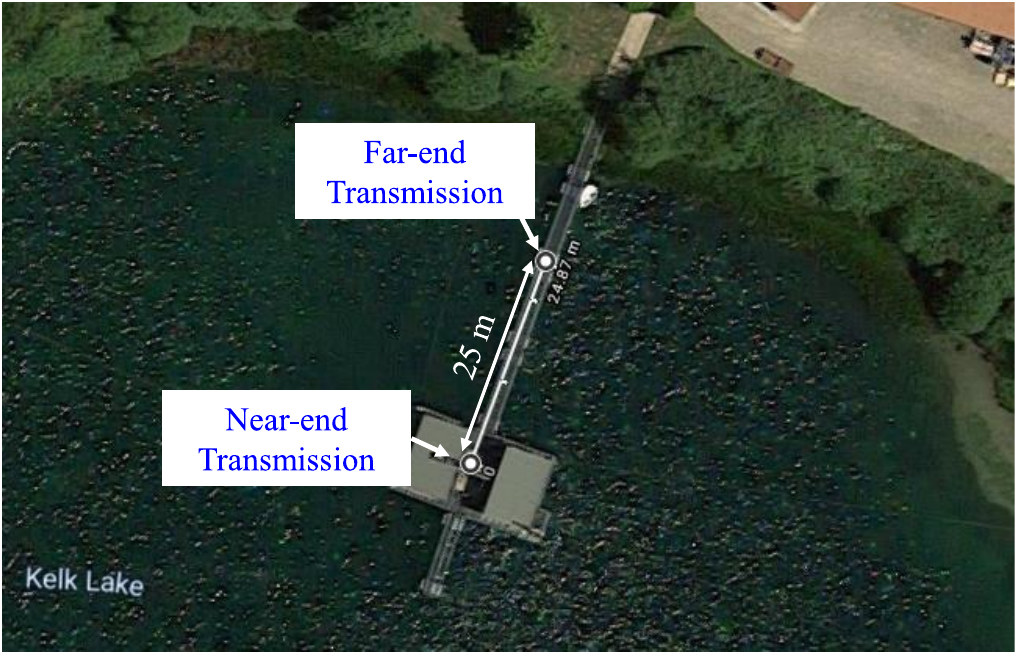}      
  \caption{\label{Fig:FD_exp_neptune} Setup of the FD and HD experiments in the Kelk lake.}
\end{figure}

\subsubsection{Experimental setup}\label{subsec:exp_setup}
Experiments were conducted in the Kelk lake in East Yorkshire. The experimental setup is shown in Fig.~\ref{Fig:FD_exp_neptune}. During the experiment, the near-end transducer was placed at 4~m depth and the far-end transducer was placed at approximately 1~m depth. The maximum depth of the site is around 7~m. The distance between the transducers was around~25~m. 

Both the near-end and far-end signals of 60~s length are transmitted at $f_c = 12$~kHz carrier frequency with $1$~kHz bandwidth. 
\textcolor{black}{The near-end source level is 150~dB re~$\mu$Pa~$@1$~m. The far-end source level is 134~dB re $\mu$Pa $@1$~m.}
The far-end SNR was around $19$~dB. For the FD experiment, the SI signal to noise ratio was around 60~dB. \textcolor{black}{The SIR in this experiment is about -41~dB.}

\subsubsection{SI and far-end channel estimation}
In this scenario, the time-variation of the SI channel is not significant, therefore the SRLSd adaptive filter is capable of providing good near-end SI estimation performance in all turbo iterations; therefore, the SRLSd algorithm is used for estimation of both the SI and far-end channels. 
For the SI channel estimation, the filter parameters are: $L_\text{SI} = 50$ taps, which corresponds to a  delay spread of $50$~ms (at the baseband sampling rate $f_d$ = 1~kHz), the sliding window length in the first iteration is $M = 1201$ and, in the further iterations, $M = 1001$. 
In this scenario, the delay spread of the far-end channel is much longer (about 30~ms) compared to that in the first lake experiment (5~ms).
For the far-end channel estimation, the filter parameters are: the filter length $L_\mathrm{f} = 30$ taps (30~ms) and the sliding window length in the first iteration is $M = 1001$ and, in the further iterations, $M = 801$.

\begin{figure}
\centering
 \begin{subfigure}{0.49\textwidth}
   \includegraphics[width=\textwidth]{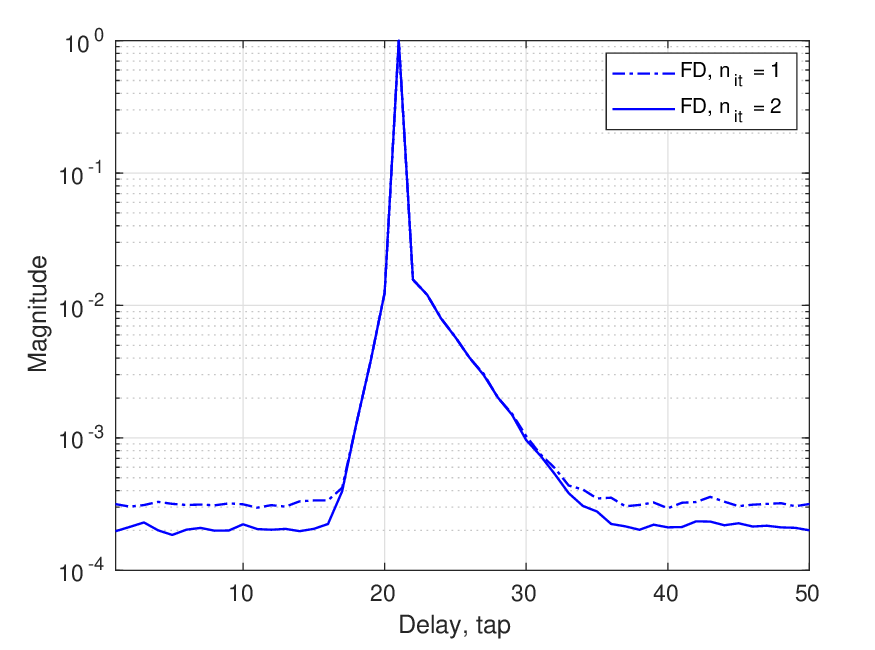}\caption{\textcolor{black}{Near-end channel estimates.}}  
  \end{subfigure}  
  \begin{subfigure}{0.49\textwidth}
   \includegraphics[width=\textwidth]{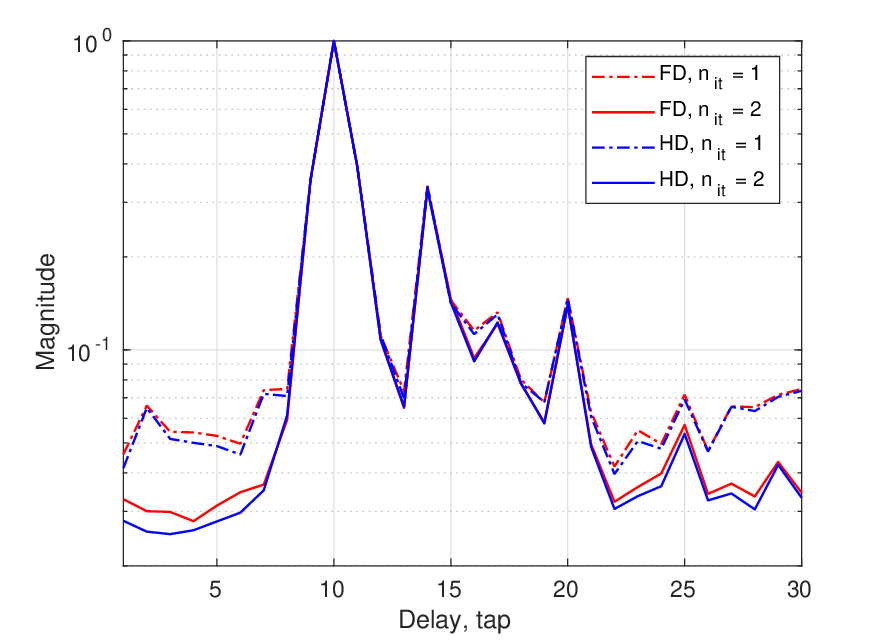} \caption{Far-end channel estimates}   
  \end{subfigure}  
  \caption{\label{Fig:far_end_est_neptune} \textcolor{black}{Near-end and} far-end channel estimates in the Kelk lake experiments with a rate 1/3 convolutional code at an SNR of 3~dB.}
\end{figure}

\begin{figure}
\centering
 \begin{subfigure}{0.49\textwidth}
   \includegraphics[width=\textwidth]{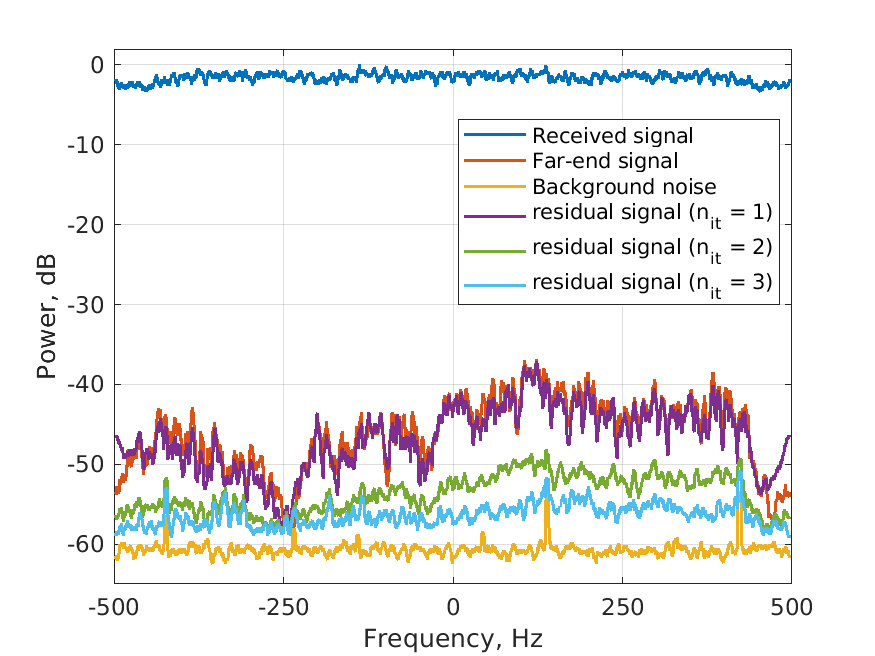}\caption{\textcolor{black}{Rake combiner.}}  
  \end{subfigure}  
  \begin{subfigure}{0.49\textwidth}
   \includegraphics[width=\textwidth]{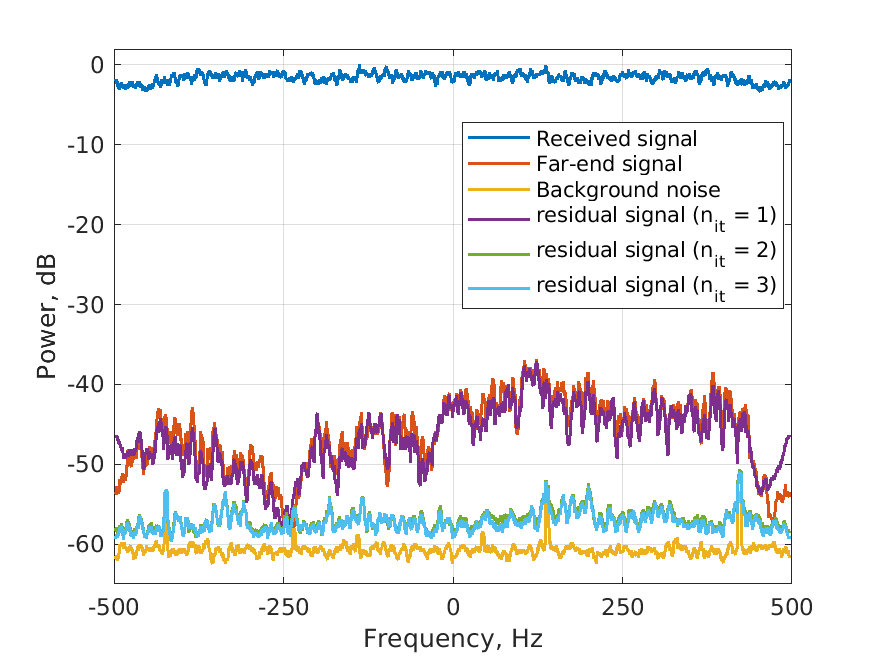} \caption{\color{black}{Rake-IC combiner}}   
  \end{subfigure}  
  \caption{\label{Fig:spectrum} \textcolor{black}{Power spectra of the baseband received signals in HD and FD experiments with a rate 1/3 convolutional code and the residual SI after SIC in three iterations. The power spectra are normalized
with respect to the maximum of the received signal spectrum.} }
\end{figure}

Fig.~\ref{Fig:far_end_est_neptune} shows the far-end channel estimates in the HD and FD experiments. The channel impulse responses are normalized with respect to the tap with the highest magnitude and plotted in log scale. Results are obtained for rate 1/3 convolutional code when the far-end SNR is about 3~dB. It can be seen that the accuracy of the channel estimates is significantly improved at the second iteration in both the HD and FD experiments. Another conclusion is that the SI signal is cancelled quite well in the FD experiment as the channel estimates obtained in HD and FD experiments are very similar.

\textcolor{black}{Fig.~\ref{Fig:spectrum} shows the power spectra of the baseband received signal, the baseband far-end signal, the background noise level and the residual signals after each iteration of SIC with Rake and Rake-IC combiners. The baseband received signal contains the near-end SI signal, the far-end signal and the background noise. The baseband residual signal is computed by subtracting the estimates of the baseband far-end signal $\hat{f}(i)$ from the SI canceller output $e(i)$. By using the Rake-IC combiner, lower levels of residual SI signals are achieved. Furthermore, the Rake-IC combiner provides a more flat spectrum of the residual SI signal. This demonstrates its ability of eliminating the multipath interference. }

\subsubsection{Demodulation performance}
Complex-valued Gaussian random noise with variance $\sigma_n^2$ is added to the baseband received signal in the HD and FD experiments as described in subsection~\ref{subsec:campus_lake} to obtain the demodulation performance at various far-end SNRs.

\begin{figure}
\begin{subfigure}{0.5\textwidth}
\centering
   \includegraphics[width=\textwidth]{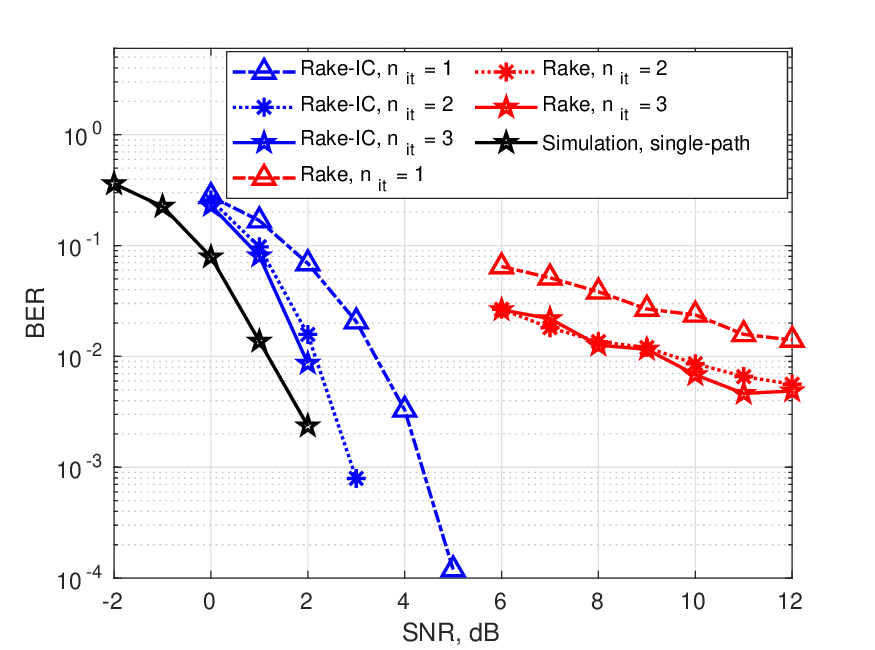}      
  \caption{Performance of \textcolor{black}{the conventional Rake and Rake-IC} combiners in HD experiments.\label{Fig:BER_1by3_neptune_Rx_comparison}}
\end{subfigure}
\begin{subfigure}{0.49\textwidth}
\centering
   \includegraphics[width=\textwidth]{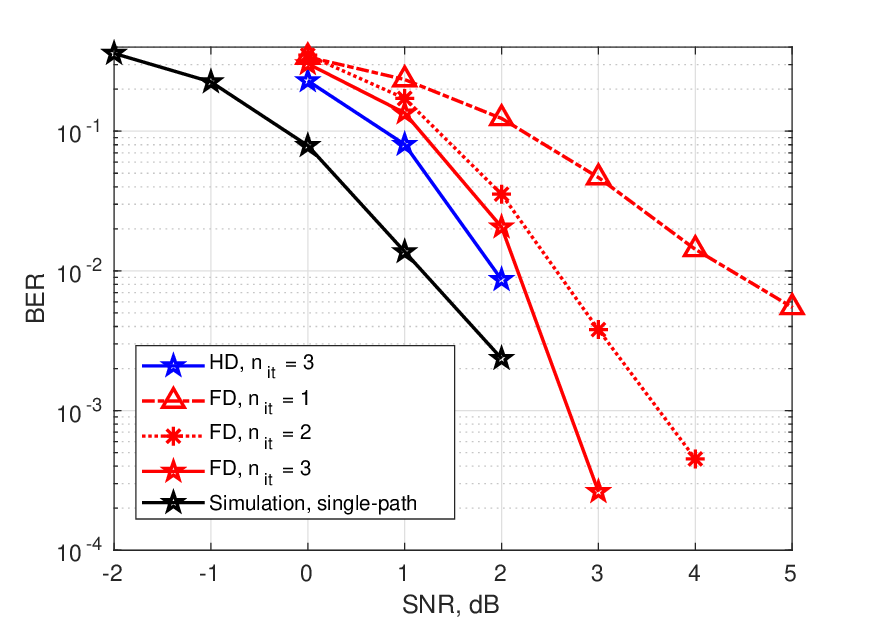}      
  \caption{Performance of the Rake-IC combiner in HD and FD experiments.\label{Fig:BER_1by3_neptune_Rake_IC} }
\end{subfigure}
 \caption{BER performance of the Rake combiners in the Kelk lake experiments with rate-1/3 convolutional code. \label{Fig:BER_1by3_neptune}}
\end{figure}

The demodulation performance of the two Rake combiners in HD experiments with rate-1/3 convolutional code are shown in Fig.~\ref{Fig:BER_1by3_neptune_Rx_comparison}. It can be seen that the performance of the Rake-IC combiner is very close to the single-path benchmark; the difference is within~1~dB. Even with single iteration, the SNR loss compared to the benchmark is only about 2~dB. On the other hand, in this scenario, the performance of the conventional Rake combiner significantly degrades due to the multipath interference in the far-end channel. An SNR loss of about 10~dB can be observed when comparing with the Rake-IC combiner.

As the performance of the conventional Rake combiner is already limited in the HD experiments, only the Rake-IC combiner is considered in the FD experiments. The demodulation performance of the Rake-IC combiner with rate-1/3 convolutional code is shown in Fig.~\ref{Fig:BER_1by3_neptune_Rake_IC}. It can be seen that the SNR loss in the FD transmission mode compared to the HD mode is less than~0.5~dB. 

\begin{figure}
\centering
   \includegraphics[width=0.5\textwidth]{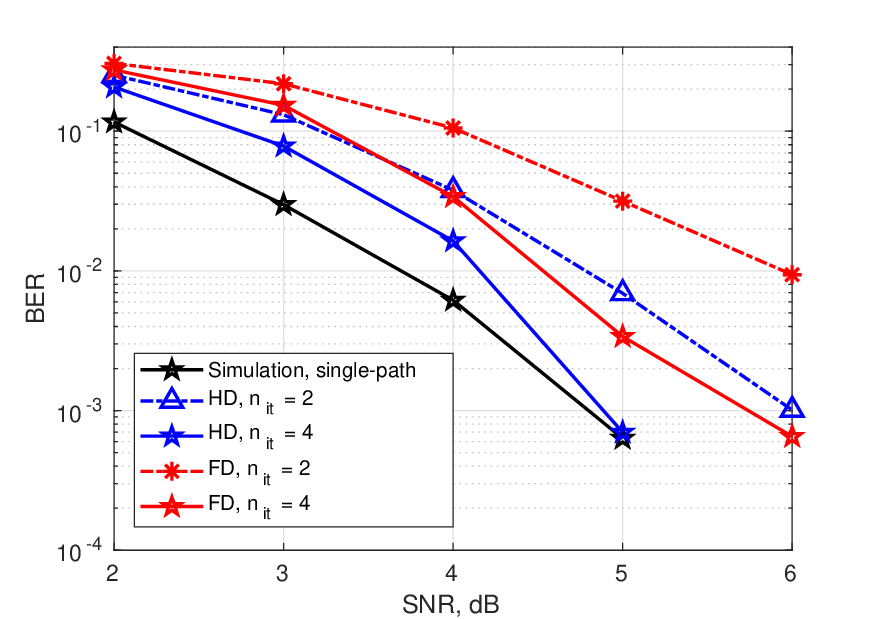}      
  \caption{\label{Fig:BER_MRC_1by2_neptune} BER performance of the Rake-IC combiner in the Kelk lake experiments with rate-1/2 convolutional code.}
\end{figure}

We then investigate the demodulation performance of the Rake-IC combiner with rate-1/2 convolutional code. 
As the code rate increases, more iterations are required to achieve the best BER performance.
As can be seen in Fig.~\ref{Fig:BER_MRC_1by2_neptune}, the performance in the HD experiment approaches the single-path benchmark in four iterations. In the FD experiments, the performance is still comparable with that in the HD experiments and with the single-path HD benchmark; the loss to the benchmark is about~1~dB at low BERs. 

\begin{figure}
\centering
   \includegraphics[width=0.5\textwidth]{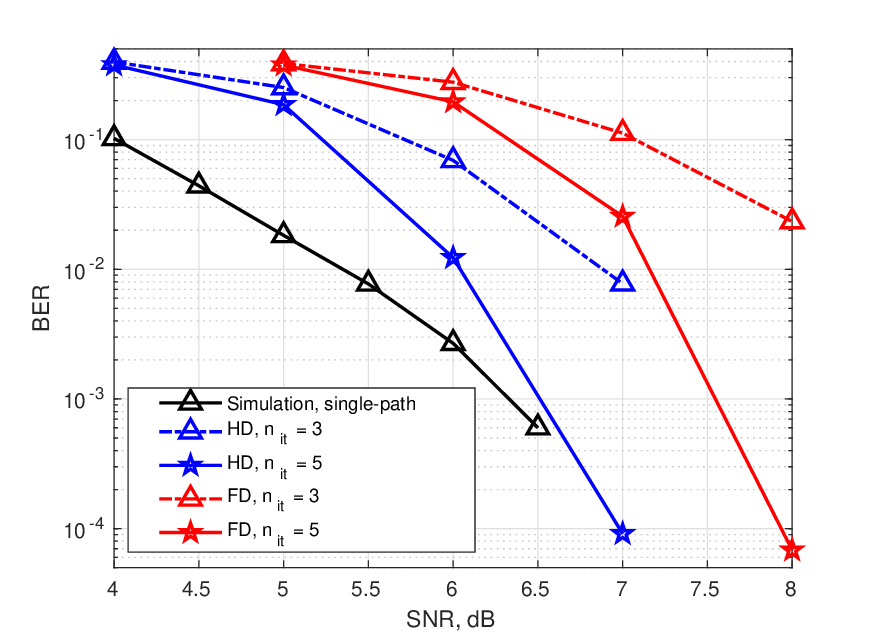}      
  \caption{\label{Fig:BER_MRC_2by3_neptune} BER performance of the Rake-IC combiner in the Kelk lake experiments with rate-2/3 convolutional code.}
\end{figure}

Finally, the demodulation performance of the Rake-IC combiner with rate-2/3 convolutional code is shown in Fig~\ref{Fig:BER_MRC_2by3_neptune}.
Now, five iterations are required to approach the single-path benchmark in the HD experiments. The performance gap between the FD and HD transmission, after five iterations, is about 1~dB at low BERs. 

Based on the results shown in Fig.~\ref{Fig:BER_1by3_neptune}, \ref{Fig:BER_MRC_1by2_neptune} and Fig.~\ref{Fig:BER_MRC_2by3_neptune}, it can be concluded that the proposed FD UWA system with the Rake-IC combiner is capable of providing similar demodulation performance as that of the HD system and the number of turbo iterations should be increased for higher data rates.

\textcolor{black}{A high residual SI level is the key issue limiting the performance of FD UWA systems. Theoretically, the SI can be cancelled to the noise level with joint near-end and far-end channel estimation. However, this remains a challenging task due to the fast variation of the SI channels and the required SI channel estimation accuracy. For example, FD lake experiments are conducted in~\cite{towliat2024joint} at two SIR levels. For an SIR of -20~dB, 22~dB of SIC is achieved with 7 hydrophones, resulting in a residual SI level 8~dB higher than the noise level. For an SIR of -15~dB, 23.5~dB SIC is achieved with 7 hydrophones, resulting in a residual SI level 10~dB higher than the noise level. For both experiments, the performance gap in the HD and FD systems is more than 7~dB. }

\textcolor{black}{ In this work, the near-end SI channel and far-end channel are estimated separately and the residual SI level is reduced in turbo iterations. According to the results presented in Section~\ref{sec:lake_exp}, the performance gap between the HD and FD systems is about 2~dB for an SIR of -51~dB. For an SIR of -41~dB, the performance gap between the HD and FD systems is about 1~dB. Therefore, it can be concluded that the proposed FD UWA system demonstrates a better performance compared with the state-of-the-art.  }

\section{Conclusions}\label{sec:conclude}
In this paper, \textcolor{black}{we consider FD UWA communication with a two-element transducer.}
An FD UWA communication system with turbo iterations and the adaptive Rake-IC combiner has been proposed. By performing the near-end SIC and far-end data demodulation in an iterative way, the influence of the far-end signal on the SIC performance has been reduced. 
The use of the adaptive Rake-IC combiner improves the demodulation performance in time-varying multipath channels compared to the conventional Rake combiner. The performance of the FD UWA system has been investigated and compared with the HD counterpart in lake experiments with different code rates. As demonstrated by the experimental results, the proposed FD UWA system with the Rake-IC combiner achieves good demodulation performance in all the experiments.

\section*{\textcolor{black}{Appendix}}
\textcolor{black}{In this section, a general description of the SRLSd and HSRLS-L-DCD adaptive filters used for SI channel estimation is given.}


\textcolor{black}{For the classical SRLS adaptive filter, the channel estimate $\mathbf{\hat{h}}(i)$ can be seen as an average of $\mathbf{h}(i)$ over the past $M$ time instants, where  $M$ is the sliding window length. If the SI channel is time-invariant, $\mathbf{\hat{h}}(i)$ can be an accurate estimate of $\mathbf{h}(i)$. However, for a time-varying channel, $\mathbf{\hat{h}}(i)$ is not an accurate estimate of $\mathbf{h}(i)$ due to the time average.
If we assume that the channel response varies linearly in the vicinity of $i$, then its average over the rectangular window is equal to $\mathbf{h}(i-M/2)$. In such a case, $\mathbf{\hat{h}}(i)$ is a more accurate estimate of $\mathbf{h}(i-M/2)$ than $\mathbf{h}(i)$.}

\textcolor{black}{For the SRLSd algorithm, the channel estimate is obtained in the same way as in the classical SRLS algorithm. 
However, when computing the error signal of the adaptive filter output, a delay of $T = M/2$ is introduced to both the input signal and the desired signal.}

\textcolor{black}{In~\cite{shen2022bem}, a low-complexity interpolation adaptive filter is proposed based on basis expansion model (BEM) approach for estimation of fast time-varying channels.
We assume that within a time interval $[i-M_o, \ i+M_o]$ centred at the time instant $i$, the time-varying response can be accurately approximated by $(P+1)$ basis functions $\phi_p(k)$: 
\begin{equation}
\mathbf{h}(i+k) = \sum_{p = 0}^{P} \mathbf{c}_p(i) \phi_p(k)  ,  \ \ k = -M_o, \ldots, M_o ,
\label{Eq:True_Channel}
\end{equation}
where $M_o = (M-1)/2$. The $L \times 1$ vectors of expansion coefficients $\mathbf{c}_p(i)$  should be estimated, where $L$ is the length of the impulse response. 
By using Legendre polynomials as the basis functions, we arrive at the SRLS-L adaptive filter.}

\textcolor{black}{The SRLS-L algorithm is summarized in Table~\ref{Tab:SRLS-L}, where $\mathbf{\Phi}_p = \mathrm{diag} \{ \phi_p(M_o), \ldots, \phi_p(-M_o) \}$ is an $M \times M$ diagonal matrix,  $\mathbf{S}(i) = [\mathbf{s}(i), \ldots, \mathbf{s}(i-M+1)]^T$ is the $M\times L$ regressor matrix, $\mathbf{x}(i) = [x(i), x(i-1), \ldots, x(i-M+1)]^T$ is an $M\times 1$ desired signal vector, $\mathbf{W} = \mathrm{diag}\{ w(M_0), \ldots, w(-M_0)\}$ is an $M \times M$ diagonal matrix, where the diagonal elements (weights) form a non-negative symmetric bell-shaped window, $\mathbf{R}(i) =  \mathbf{s}(i)\mathbf{s}^H(i)$, $\varphi_p(k) = \sqrt{w(k)}\phi_p(k)$ and $\varepsilon$ is a regularization parameter introduced to avoid the numerical instablity due to the matrix inversion.}

\begin{table}[t]

\scriptsize
\centering
\caption{\textcolor{black}{SRLS-L algorithm}\label{Tab:SRLS-L}}
\begin{tabular}{ll}

\hline
\textbf{Step} & \textbf{Equation} \\ \hline
     & for $i > 0$, repeat: \\
  1  & Generate vectors $\mathbf{b}_{p}(i+M_o) = \mathbf{S}^T(i+M_o) \mathbf{W}\mathbf{\Phi}_p \mathbf{x}^*(i+M_o),$\\
 &$p = 0, \ldots, P$ \\
  2  & Compute matrices $\mathcal{R}_{p,q}(i) = \sum_{k=-M_o}^{M_o} \varphi_p(k)  \mathbf{R}(i+k) \varphi_q(k), $\\
  &$p,q  = 0, \ldots, P$ \\
  3  & Generate the matrix $\mathcal{R}(i)$ and vector $\mathbf{b}(i)$ \\
  4  & Find a solution $\hat{\mathbf{c}}$ to the system $\left[\mathcal{R}(i) + \varepsilon \mathbf{I}_{(P+1)L} \right] \mathbf{c}(i) = \mathbf{b}(i)$\\
  5  & Compute the estimate $\hat{\mathbf{h}}(i) = \sum_{p = 0}^{P} \hat{\mathbf{c}}_p(i) \phi_p(0) $  \\
      \hline
\end{tabular}
\end{table}

\begin{table}[]
\scriptsize
\centering
\caption{\textcolor{black}{H$\ell_1$-DCD algorithm}\label{Tab:Hl1-DCD}}
\begin{tabular}{cl}
\hline
 & Input parameters: $M_0$, $P$, $L$,  $\gamma$, $\mu_d$, $\mu_w$ \\
 & Output: $\mathbf{\hat{c}}$, $\mathbf{\tilde{w}}$\\
Step & Initialization:  $I = \emptyset$, $\mathcal{R} = \mathcal{R}(i)$, $\mathbf{c} = \mathbf{0}$, $\mathbf{b} = \mathbf{r}(i)$, $\mathbf{\tilde{w}} = \mathbf{1}_L$ \\ \hline
     & for $i > 0$, repeat: \\
  1  & $\tau = \mathrm{max}_k|b_k|$ \\
  2  & Remove the $t$th element from $I$ ($I \leftarrow I \,\backslash\,t$), if \\
     & \quad $t = \mathrm{arg}\,\mathrm{min}_{k\in I}\frac{1}{2}|c_k|^2\mathcal{R}_{k,k} + \Re\{c_k^*b_k\} - \tau \tilde{w}_k|c_k|$ \\
     & and $\frac{1}{2}|c_k|^2\mathcal{R}_{k,k} + \Re\{c_k^*b_k\} - \tau \tilde{w}_k|c_k| < 0$    \\
  3  & If the $t$th element is removed, then update:  \\
     & \quad $\mathbf{b} = \mathbf{b} + c_t\mathcal{R}^{(t)}$ \\
  4  & Include the $t$th element into the support ($I \leftarrow I \,\cup\,t$), if \\
     & \quad $t = \mathrm{arg}\,\mathrm{max}_{k\in I}\dfrac{(|b_k| -\tau \tilde{w}_k)^2}{\mathcal{R}_{k,k}}$  and $|b_t| > \tau \tilde{w}_t$ \\
  5  & Update the regularization parameter: $\tau = \gamma \tau$ \\
  6 & Approximately solve the LS-$\ell_1$ optimization on the support $I$ \\
     & using the leading $\ell_1$-DCD algorithm~\cite{zakharov2008low} \\
  7 & Debiasing according to $I = \{k: |c_k| > \mu_d\mathrm{max}_k\{|c_k|\}\}$\\
  8 & Reweighting according to $\mathbf{\tilde{w}}(i) = (1 - \mu_w)\mathbf{\tilde{w}}(i-1) + \mu_w \bar{\mathbf{w}}$ \\
      \hline
\end{tabular}
\end{table}

\textcolor{black}{For fast-varying channels with a large delay spread, the minimum sliding window length required is significantly increased when high orders of the basis functions are used. In practice, there is sparsity in the expansion coefficients. By exploiting the sparsity, the sliding window length can be reduced, which in turn will improve the tracking performance of the SRLS-L algorithm.}

\textcolor{black}{By solving the sparse recovery problem using the homotopy principle and DCD iterations, we arrive at the HSRLS-L-DCD adaptive filter, which uses the H$\ell_1$-DCD algorithm to find the solution $\hat{\mathbf{c}}$ in Step 4 of the SRLS-L algorithm~\cite{zakharov2012homotopy}. The H$\ell_1$-DCD algorithm is summarized in Table~\ref{Tab:Hl1-DCD}, where $\gamma < 1$ is a positive factor used for reducing the regularization parameter $\tau$, $\mu_d$ is a predefined parameter between zero and one for removing low-magnitude components from the solution and $\mu_w \in (0,1]$ is a parameter which defines the update rate. }

\section{Acknowledgement}
The work of L. Shen, B. Henson and  Y. Zakharov was supported in part by the U.K. EPSRC through Grants EP/R003297/1 and EP/V009591/1. Part of the materials in this paper was presented at the Sixth Underwater Communications and Networking Conference (Ucomms) in Lerici, Italy.

\bibliography{FD_exp}
\bibliographystyle{IEEEtran}
\end{document}